\documentclass{PoS}

\usepackage{epsfig}

\newcommand{\be}{\begin{equation}}
\newcommand{\ee}{\end{equation}}
\newcommand{\ba}{\begin{eqnarray}}
\newcommand{\ea}{\end{eqnarray}}
\newcommand{\bi}{\begin{itemize}}
\newcommand{\ei}{\end{itemize}}

\newcommand{\nn}{\nonumber \\}
\newcommand{\dg}{^\dagger}

\newcommand{\eq}{Eq.~}

\newcommand{\fig}{Fig.~}
\newcommand{\tab}{Tab.~}
\newcommand{\la}{\label}

\newcommand{\txts}{\textstyle}

\newcommand{\Lmsbar}{\mathop{\Lambda_{\overline{\rm MS}}}}

\newcommand{\Fps}{\mathop{F_{\rm ps}}}
\newcommand{\fps}{\mathop{f_{\rm ps}}}

\newcommand{\mps}{\mathop{M_{\rm ps}}}
\newcommand{\Nf}{{N_{\rm f}}}
\newcommand{\mref}{\mathop{m_{\rm ref}}}
\newcommand{\Lmax}{\mathop{L_{\rm max}}}

\newcommand{\hQ}{\hat Q}
\newcommand{\hQa}{\hat Q_A}
\newcommand{\Mee}{M_{\rm ee}}
\newcommand{\Meo}{M_{\rm eo}}
\newcommand{\Moe}{M_{\rm oe}}
\newcommand{\Moo}{M_{\rm oo}}
\newcommand{\Mmee}{M^{-1}_{\rm ee}}
\newcommand{\Mmoo}{M^{-1}_{\rm oo}}
\newcommand{\gf}{\gamma_5}

\title{Trajectory Length and Autocorrelation Times -- 
       $N_f=2$ simulations in the Schr\"odinger functional}

\ShortTitle{Trajectory Length and Autocorrelation Times}

\author{\speaker{Harvey~Meyer}\\
        Deutsches Elektronen-Synchrotron DESY\\
        Platanenallee 6, D-15738 Zeuthen\\
        E-mail: \email{harvey.meyer@desy.de}}

\author{Oliver Witzel\\
Humboldt-Universit\"at, Institut f\"ur Physik \\
Newtonstrasse 15, D-12489 Berlin}

\abstract{A status report is presented on the large-volume simulations
in the Schr\"odinger functional with two flavours of O($a$) improved
Wilson quarks performed by the ALPHA collaboration.
The physics goal is to set the scale for the computation of the 
fundamental parameters of QCD.
In this talk the emphasis is on aspects of the Hybrid Monte-Carlo algorithm,
which we use with (symmetric) even-odd and Hasenbusch preconditioning. 
We study the dependence of aucorrelation times on the trajectory length.
The latter is found to be significant for fermionic correlators,
the trajectories longer than unity performing better than the shorter ones.
\vspace{0.7cm}\\  DESY 06-161 \\ HU-EP-06/28 \\ SFB/CPP-06-42}

\FullConference{XXIV International Symposium on Lattice Field Theory\\
		 July 23-28 2006\\
		 Tucson Arizona, US}

\begin{document}

\section{Motivation}
Recently the $\Lambda$-parameter of two-flavour QCD has been computed 
by the {ALPHA} collaboration in the Schr\"odinger functional scheme in units 
of a low-energy scale $\Lmax$~\cite{alpha2}. The latter scale is defined implicitly
by the renormalized coupling in that scheme taking a particular value,
$\bar g^2_{\rm SF}(\mu=1/L)=4.61$. The determined quantity 
$\Lambda_{\rm SF}\Lmax$ is a continuum, universal result.
The connection between $\Lambda_{\rm SF}$ and $\Lmsbar$ being known 
exactly~\cite{sint-sommer}, what remains to be done, for the result to be
phenomenologically useful, is to trade $\Lmax$ 
for an experimentally accessible low-energy scale. 
This was done provisionnally~\cite{alpha2} using 
the chirally extrapolated Sommer reference scale $r_0$~\cite{r0qcdsf,alpha2}, 
however choosing a more directly accessible physical quantity 
will reduce the systematic uncertainty. 
Our present goal is to use the kaon decay constant $F_K$ as low-energy scale.
This will also allow us to reduce the overall uncertainty on our recent 
determination of the strange quark mass~\cite{m_s}.

More specifically, a possible strategy is to compute $\Lmax \Fps$ at a tuned 
quark mass such that $\frac{\mps}{\Fps}(\mref,\mref)= \frac{M_K}{F_K}|_{\rm exp}$.
Neglecting the quenching of a third quark of mass $\mref$, 
SU(3) chiral perturbation theory then connects
$\Fps(\mref,\mref,\mref)$ to $\Fps(m_u,m_d,m_s)$, which is 
equated to its experimental value of $113{\rm MeV}$. 
At one-loop level~\cite{gl3}, with 
$\mu_P =  \textstyle{\frac{M_P^2}{32\pi^2F_0^2}}~\log\{M_P^2/\mu^2 \}$  
and $F_0=88$MeV we have 
\ba
\frac{\mref}{(m_\ell+m_s)/2} &=& 1+ \txts{\frac{3}{2}}\mu_\pi - \txts{\frac{11}{3}}\mu_K
                                +\txts{\frac{13}{6}}\mu_\eta 
                                + \frac{16}{F_0^2}[L_4(\mu)-L_6(\mu)]\cdot[M_K^2-M_\pi^2]
\la{eq:mref}\\
    & = &1~ + ~ 0.10 ~+~ 0.47\cdot 10^3[L_4(M_\rho)-L_6(M_\rho)], \nn
\frac{\Fps(\mref,\mref,\mref)}{F_K(m_\ell,m_\ell,m_s)}
&=& 1+
\textstyle{\frac{3}{4}}(\mu_\pi - 2\mu_K + \mu_\eta)  
+\textstyle{\frac{4}{F_0^2}}\cdot L_4(\mu)\cdot[M_K^2-M_\pi^2]
\la{eq:FpsFK}\\
&=&1 ~+~ 0.051 ~+~ 0.12 \cdot 10^3L_4(M_\rho),\nonumber
\ea
where phenomenological studies point to $10^3|L_4|\leq 1$ and 
$0.2\leq10^3[L_4(M_\rho)-L_6(M_\rho)]\leq0.4$~\cite{bijnens,meissner}.
These ranges can be narrowed down by lattice simulations.
Using the tree-level relation $4M_K^2-3M_\eta^2=M_\pi^2$, it can be verified that 
the dependence on $\mu$ cancels in \eq (\ref{eq:mref}) and (\ref{eq:FpsFK}).

Although the original purpose of the Schr\"odinger functional was to 
provide the setting for the implementation of a finite-volume renormalization 
scheme, it was shown~\cite{guagnelli}  in the quenched approximation
that it also provides a competitive tool 
for spectrum and matrix element calculations. Roughly speaking,
the Dirichlet boundary conditions in the time directions are now exploited 
in the same way as in the  source method that was 
used in the early days of glueball spectroscopy~\cite{source}.

\section{Simulations at $\mref$}
We use the $\Nf=2$ O($a$) improved Wilson formulation with plaquette gauge action $S_{\rm g}$.
Our simulations are performed with
the Hybrid Monte-Carlo algorithm with (symmetric) even-odd~\cite{jliu}
and Hasenbusch preconditioning~\cite{hasenb}. The action thus reads
$S= S_{\rm pf} ~+~ S_{\det}  ~+~ S_{\rm g}$, with
\be
S_{\rm pf}=\phi_0\dg \frac{1}{\hQ\hQ\dg+\rho_0^2\Moo^{-2}}\phi_0 +
    \sum_{k=1}^{n-1}     \phi_k\dg \left(\frac{1}{\sigma_k^2}+
    \frac{1}{\hQa^2+\rho_k^2}\right)\phi_k
\ee
where the pseudofermions (PFs) `live' on the odd sites only and, 
in the notation of~\cite{jliu},
\ba 
\hQa &=& \gf (\Moo-\Moe \Mmee \Meo),\qquad \hQ=\Mmoo \hQa,  \\
S_{\det} &=&   (-2)\log\det \Mee ~+~ (-2)\log\det \Moo, 
\ea
The parameters $(\sigma_k,\rho_k)$ satisfy
\be \sigma_k^2 = \rho_{k-1}^2  - \rho_k^2,\qquad k=1,\dots n-1; 
\qquad \rho_{n-1} = 0.
\ee
The Hermitian matrix $\hQa$ is thus the Schur complement of $Q\equiv\gamma_5 D$
with respect to even-odd preconditioning. Note that $\hQ$ on the other hand is 
not Hermitian. The matrices $\Moo,\Mee$ are diagonal with respect to space indices,
and their determinants are thus represented exactly on each configuration.
All the simulations presented here are with $n=2$ unless otherwise stated.

\subsection{Recent algorithmic improvements}

\begin{figure}
\begin{center}
\psfig{file=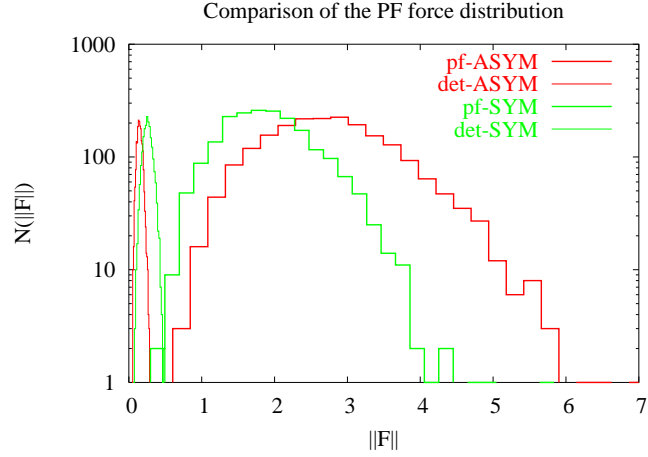,angle=0,width=9.0cm} 
\end{center}
\caption{Comparison of the pseudofermion and determinant 
forces in the time-slice $T/2$ with symmetric and asymmetric even-odd 
preconditioning on an $8^4$ lattice at $\beta=6.32$ ($n=1$ here, and the 
normalization of the force is as in~\cite{sap3}).}
\la{fig:eo-force}
\end{figure}

The advantage of the symmetric form of even-odd preconditioning was first 
demonstrated in~\cite{JLQCD} with $n=1$. See~\cite{hasenb-LAT05} for a 
study with $n=2$ (where the shift $\rho$ was however implemented on $D$ rather
than $Q$). We find that 
on a $16^4$ lattice at $\beta=5.2$, the stepsize $\delta\tau$ can be increased
from $1/28$ to $1/20$ at a constant acceptance above $90\%$, when
switching from the asymmetric to the symmetric form. This is easily 
explained by a reduction in the magnitude of the force, as illustrated on
\fig\ref{fig:eo-force} for a small system. The operator $\hQ$ is better
behaved \emph{in the ultraviolet} than $\hQa$; the improvement 
 is thus not expected to depend strongly on the size of the system.
Secondly, in the Hamiltonian and force computation, 
the equation $(\hQa^2+\rho^2)\psi=\phi$ arises, to be solved 
for $\psi$. We then proceed as follows, with $\Mmoo$ stored:
\be
\tilde\phi = \Mmoo\phi, \qquad
 (\hQ-i\rho\Mmoo)(\hQ\dg+i\rho\Mmoo)\tilde\psi = \tilde\phi,\qquad
\psi = \Mmoo \tilde \psi. \la{eq:hamilk}
\ee
We use the conjugate gradient algorithm for the  inversion involving $\hQ$,
and on the $16^4$ lattice find about a $20\%$ reduction in the
number of  iterations necessary to solve the equation with $\rho=0$
for a given target accuracy with respect to solving the equation involving $\hQa$.
Thus the symmetric variant is also a superior solver preconditioner
as compared to the asymmetric one.

It was recently suggested to introduce
a separate step-size $\delta\tau_k$ for each pseudofermion~\cite{qcdsf,urbach}
(a smaller step-size for the gauge force has been in use for a long time~\cite{Sexton:1992nu}).
With $n=2$ one then has one additional parameter to tune, and it turns out that 
the simultaneous optimization of $\rho_0$ and $\delta\tau_0/\delta\tau_1$ 
leads to significantly smaller values of $\rho_0$~\cite{urbach,alpha-unpublished} 
than if one constrains $\delta\tau_0=\delta\tau_1$. We find a
gain resulting from the introduction of the extra parameter 
of about 1.5 on a $24^3\times32$ lattice at $m\simeq\mref$,
while no observable improvement could be obtained on the $16^4$, $\beta=5.2$ lattice.
This suggests that the gain increases with the condition number of $\hQ$.

We have also explored `hybrid' integrators, where the leap-frog 
integration scheme is used for the force associated with $\phi_1$
and the Sexton-Weingarten~\cite{Sexton:1992nu} for that associated with $\phi_0$, and 
found the performance to be comparable. The motivation is to 
benefit from the robustness of the leap-frog integrator for the most
irregular force to insure stability, 
while using a higher order scheme for the smoother forces.

\section{Trajectory length dependence of autocorrelation times}

We report some results on the algorithm figure-of-merit $\nu$ introduced in~\cite{sap3},
which we generalize to any observable $O$. If $\tau_{\rm int}[O]$ is its autocorrelation time
in units of trajectories, we have 
\be
\nu[O] \equiv 10^{-3}\cdot N_{\rm eff}^{\rm solver~calls}\cdot \tau_{\rm int}[O].
\ee
By `effective number of solver calls', we mean the number of times the equation
$\hQa^2\psi=\phi$ has to be solved per trajectory, plus the number of times the
equation $(\hQa^2+\rho_0^2)\psi=\phi$ has to be solved, weighted by the number 
of solver iterations this takes relative to the former equation. It turns out
that these two contributions are roughly equal.
The values of $\nu$ for the plaquette, the effective pion mass and 
pseudoscalar decay constant in the middle time-slice are given in \tab\ref{tab:perf}.
There is certainly no sign that the algorithm's performance worsens 
as the quark mass decreases in the explored region.
The figures for the plaquette $P$ are very similar to the ones quoted 
in~\cite{luscher-LAT05}, and also in~\cite{urbach} (if one uses the same definition of
$N_{\rm eff}^{\rm solver~calls}$).

\begin{table}
\begin{center}
\begin{tabular}{c@{~~~}c@{~~~}c@{~~~}c@{~~~}c@{~~~}c@{~~~}c}
\hline\hline
$\kappa$  &  $\tau$  & integrator& $\delta\tau_1/\delta\tau_0$ & $\nu[P]$ & $\nu[\mps(T/2)]$ & $\nu[\fps(T/2)]$ \\
\hline
0.1355    & $1/2$    &  SW & 1  & $0.47(13)$& $0.67(18)$       & $1.4(5)$  \\
0.1355    & 2        &  SW & 1  & $0.80(16)$& $0.65(16)$       & $0.33(8)$ \\
0.1359    & $1/2$    &  SW & 1  & $0.36( 8)$& $0.75(21)$       & $0.45(10)$   \\
\hline
0.13605   & 2        &  LF & 5  & $0.6(2)$   & $0.25(6)$        & $0.26(7)$  \\
0.13625   & 2        &  LF & 5  & $0.55(27)$ & $0.17(6)$        & $0.16(6) $  \\
\hline\hline
\end{tabular}
\caption{The figure-of-merit $\nu[O]$ for the plaquette, the effective pseudoscalar
mass and decay constant at $\beta=5.3$
in the range of masses $m_s/2<m<m_s$ on $24^3\times32$ lattices.}
\label{tab:perf}
\end{center}
\end{table}

At $\kappa=0.1355$, two different runs with trajectory length 
$1/2$ and 2 were performed. While the plaquette actually `prefers' the shorter 
trajectory, the observables of interest (especially $\fps$)
have a smaller autocorrelation time with the longer trajectory.
Determining autocorrelation times accurately is difficult, and therefore it is
useful to look directly at the normalized autocorrelation function
$\rho(t)$. In Fig.~\ref{fig:f1fP} this function is shown for two fermionic correlators: $f_1$
which corresponds to the propagation of the $(u\bar d)$ quarks from one boundary
to the other and $f_P$, where the pseudoscalar density in the middle time slice
annihilates the states created by the boundary (see~\cite{guagnelli} for the precise definition
of these correlators). A noticeable  difference is seen at small $t$. 
To make the observation more significant, we repeat the comparison in the 
quenched theory with an equivalent quark mass, using the HMC algorithm for the 
pure gauge update, and with a much smaller spatial volume (see \fig\ref{fig:qf1fP}). 
Here the statistics is high enough to allow us to see a clear difference
between the choices $\tau=1/2,2$ and 4 for the trajectory length. The longer trajectories
are undoubtedly superior in this situation.

\begin{figure}[t]
\centerline{
\psfig{file=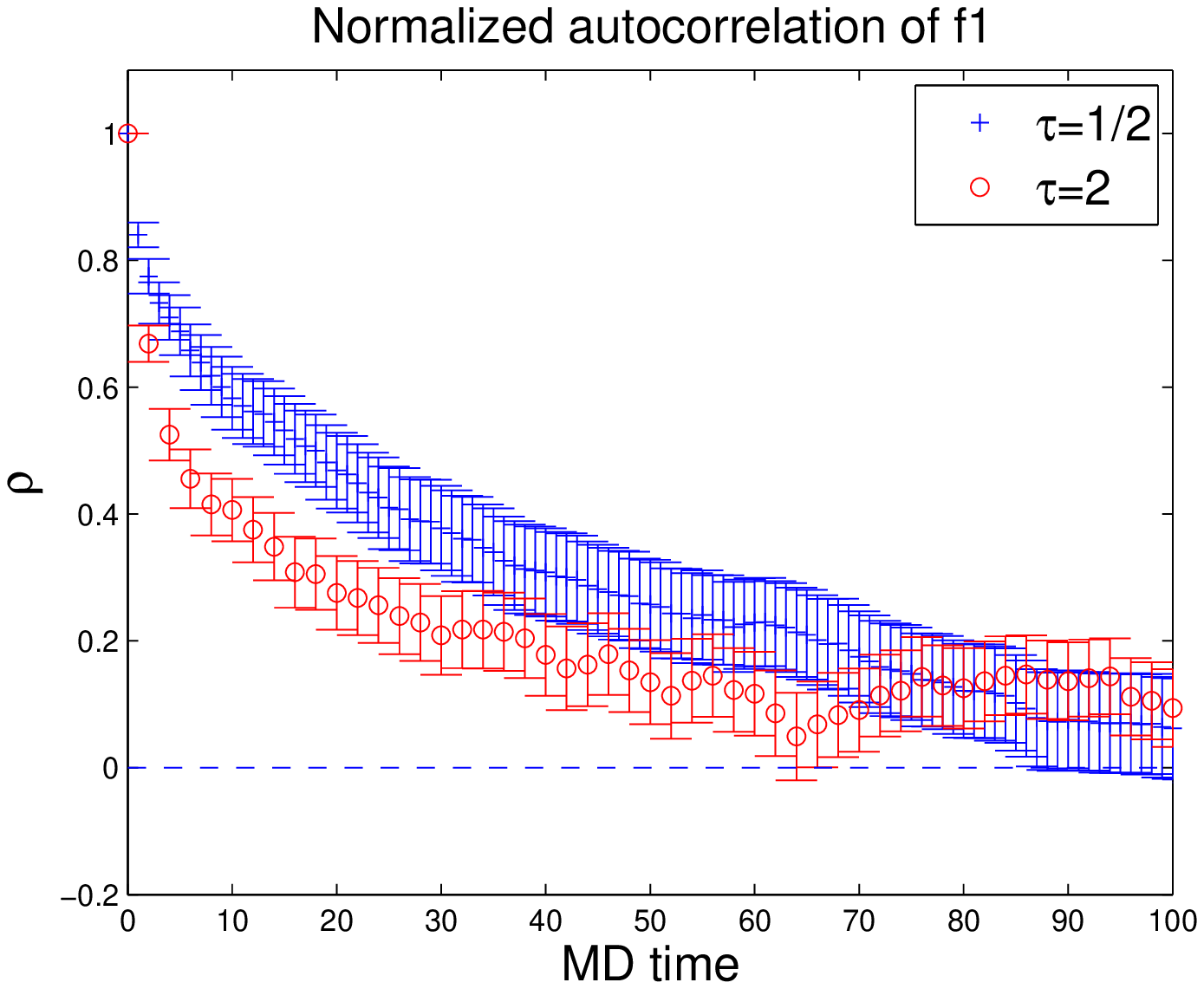,angle=0,width=7.0cm, height=5.5cm}
\psfig{file=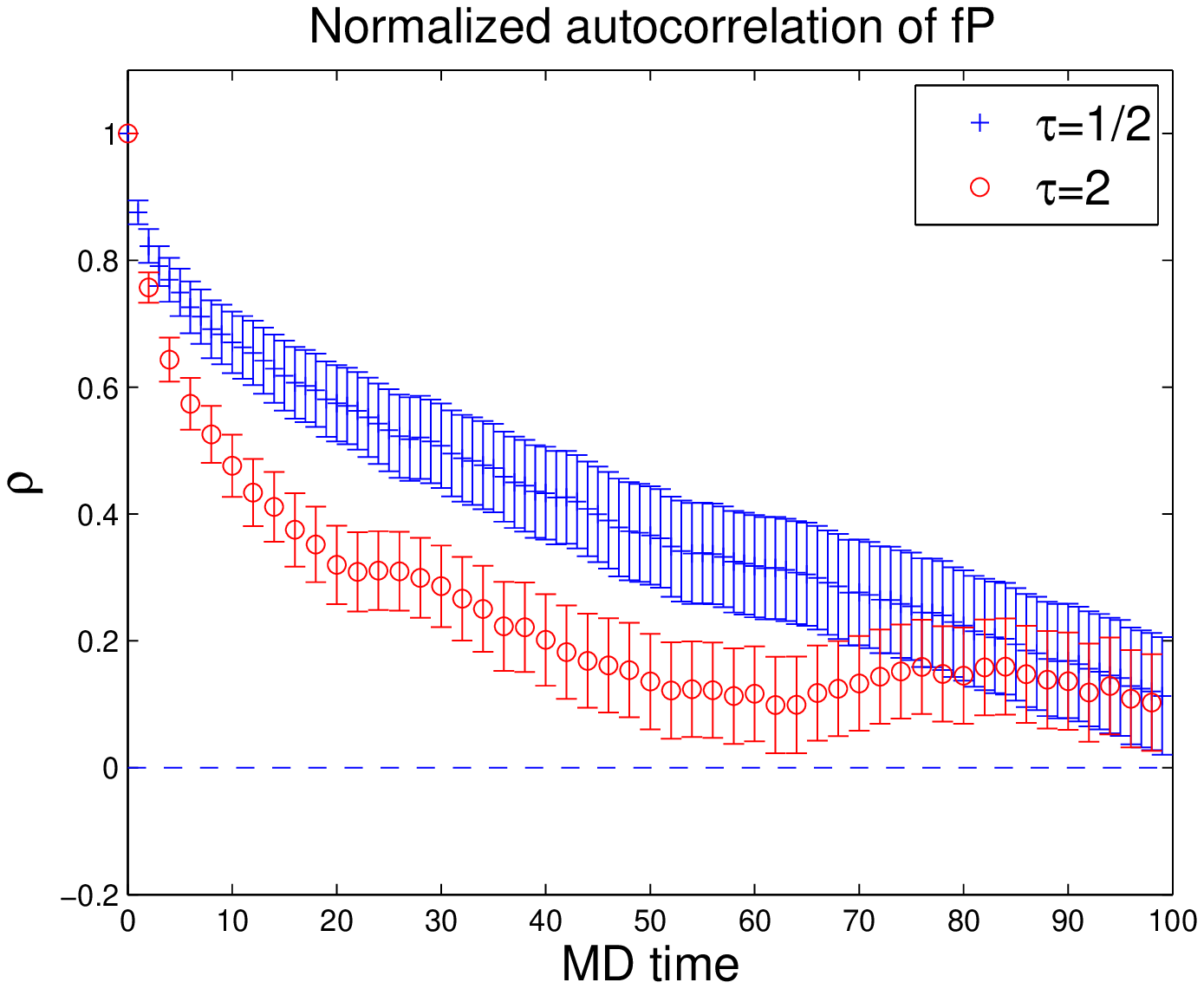,angle=0,width=7.0cm, height=5.5cm}
}
\caption{Autocorrelation function of the correlators $f_1$ and $f_{\rm P}$
($\Nf=2$, $\beta=5.3$, $\kappa=0.1355$, $V=24^3\times32$), for different trajectory lengths.}
\la{fig:f1fP}
\end{figure}

The reduced autocorrelation times observed with long trajectories should be 
reflected in faster thermalization of the same observables. We thus perform 
the following experiment. We start from an ensemble of 16 small-volume and small-quark mass
configurations with a background chromo-electric field.
The correlators $f_P$ and $f_P'$, defined by the correlation from one or the other boundary 
to the middle time-slice, are thus not equal.
We then change the SF boundary conditions so as to remove the background field, after which
$f_P=f_P'$ holds exactly in the new thermalized ensemble. \fig\ref{fig:therm} shows 
how fast $f_P-f_P'$ thermalizes to zero with two different choices of trajectory length.
The advantage of $\tau=2$ over $\tau=1/2$ is obvious in the early stages of thermalization.

\begin{figure}[t]
\centerline{
\psfig{file=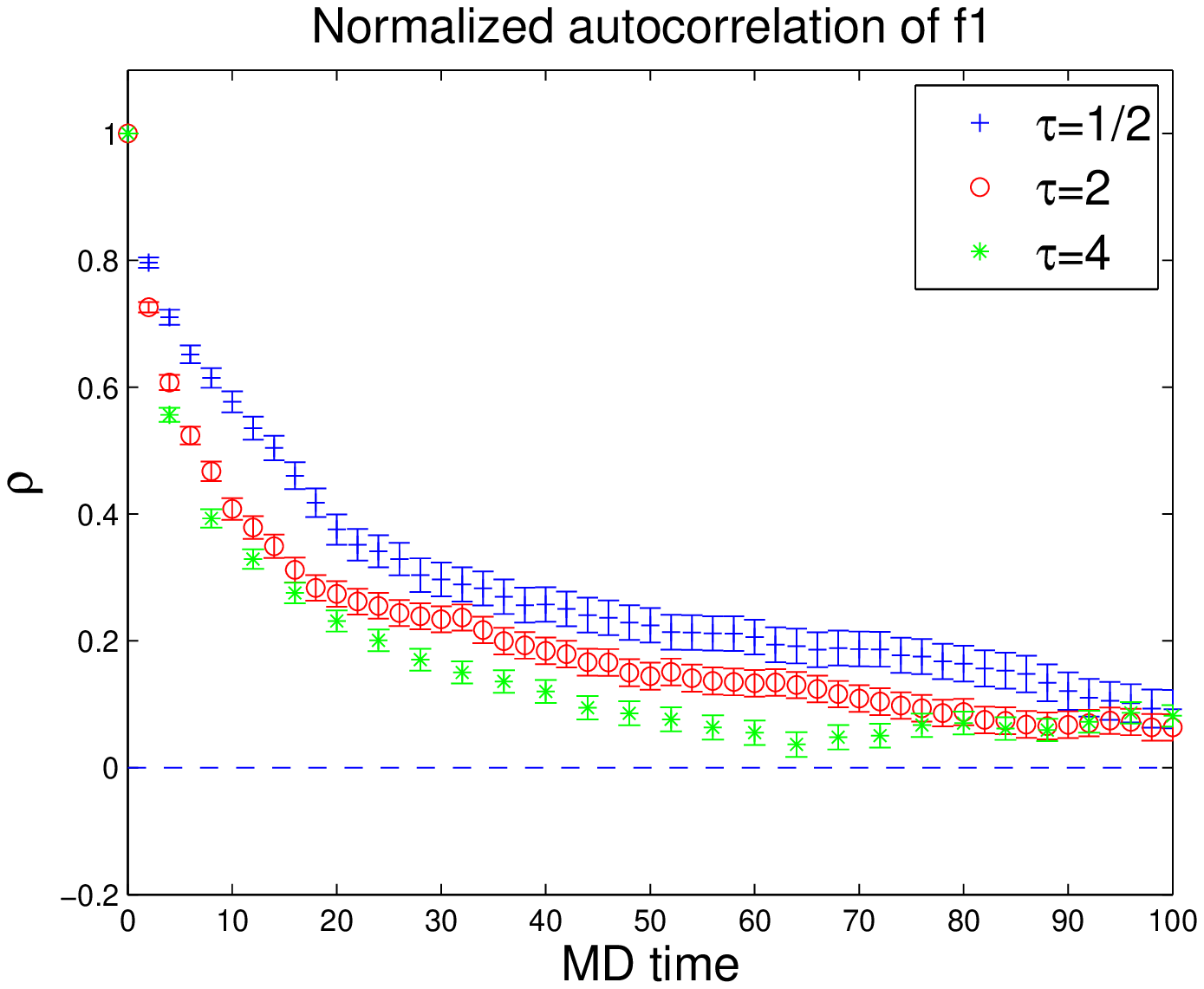,angle=0,width=7.0cm, height=5.5cm}
\psfig{file=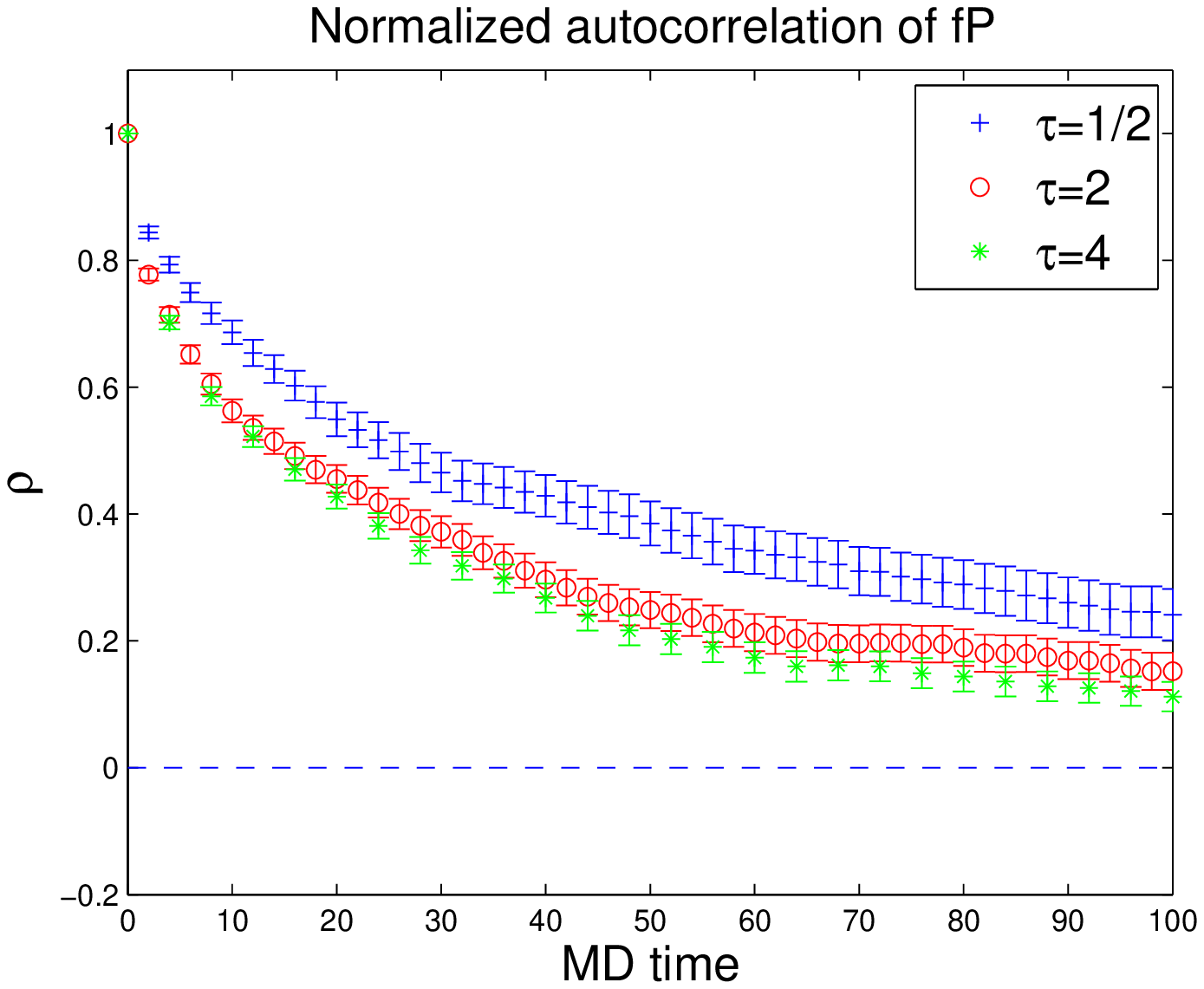,angle=0,width=7.0cm, height=5.5cm}
}
\caption{Autocorrelation function of the correlators $f_1$ and $f_{\rm P}$
in quenched QCD ($\beta=6.0$, $\kappa=0.1338$, $V=8^3\times32$), for different trajectory lengths.}
\la{fig:qf1fP}
\end{figure}

\begin{figure}[t]
\centerline{\psfig{file=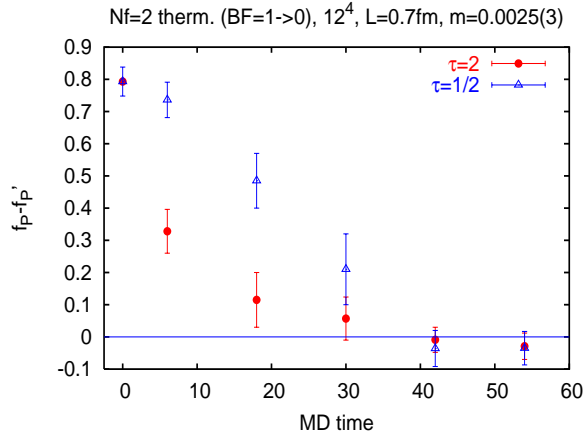,width=8.0cm, height=6cm}}
\caption{Thermalization of  $f_{\rm P}-f_{\rm P}'$,
whose expectation value vanishes. The data points are averaged over 16 
configurations  taken from an ensemble where the expectation value is non-zero.}
\la{fig:therm}
\end{figure}

\section{Conclusion}
We are performing large-volume $\Nf=2$ simulations
with O($a$) improved Wilson quarks in the Schr\"odinger 
functional at $m\simeq m_s/2$ to set the low-energy scale 
for our determination of the $\Lambda$ parameter~\cite{alpha2} 
and the strange quark mass~\cite{m_s}. Apart from the $L/a=24$ simulations
presented here, simulations at $L/a=32$ and larger $\beta$ are underway.

We have shown that the trajectory length $\tau$ is an 
important parameter of the Hybrid Monte-Carlo algorithm.
In practice we have found that $\tau\geq2$ is never a worse 
choice for long-distance observables based on fermionic correlators.
The reduction in autocorrelation time is in some cases as large as
a factor two. We have also provided evidence that 
slow-thermalization problems can be alleviated 
by choosing longer trajectories.
For more details, as well as a discussion of stability and 
reversibility issues, we refer the reader to~\cite{tlength}.

\paragraph{}
We thank Michele Della Morte, Hubert Simma, Rainer Sommer and Ulli Wolff for a 
stimulating collaboration and a careful reading of the manuscript.
We further thank DESY/NIC for computing resources on the APE machines
and the computer team for support, in particular to run on the new
apeNEXT systems.
This project is part of ongoing algorithmic development within
the SFB Transregio 9 ``Computational Particle Physics'' programme.


\begin{thebibliography}{99}

\bibitem{alpha2}
  M.~Della Morte, R.~Frezzotti, J.~Heitger, J.~Rolf, R.~Sommer and U.~Wolff
                  [ALPHA Collaboration],
  Nucl.\ Phys.\ B {\bf 713} (2005) 378
  [arXiv:hep-lat/0411025].

\bibitem{sint-sommer}
  S.~Sint and R.~Sommer,
  Nucl.\ Phys.\ B {\bf 465} (1996) 71
  [arXiv:hep-lat/9508012].

\bibitem{r0qcdsf}
M.~G\"ockeler, R.~Horsley, A.~C.~Irving, D.~Pleiter, P.~E.~L.~Rakow, G.~Schierholz and H.~Stuben
                  [QCDSF Collaboration],
Phys.\ Lett.\ B {\bf 639} (2006) 307
arXiv:hep-ph/0409312.

\bibitem{m_s}
  M.~Della Morte, R.~Hoffmann, F.~Knechtli, J.~Rolf, R.~Sommer, I.~Wetzorke and U.~Wolff
                  [ALPHA Collaboration],
  Nucl.\ Phys.\ B {\bf 729} (2005) 117
  [arXiv:hep-lat/0507035].

\bibitem{gl3}
  J.~Gasser and H.~Leutwyler,
  Nucl.\ Phys.\ B {\bf 250} (1985) 465.

\bibitem{bijnens}
  J.~Bijnens, P.~Dhonte and P.~Talavera,
  JHEP {\bf 0405}, 036 (2004)
  [arXiv:hep-ph/0404150].

\bibitem{meissner}
 T.~A.~Lahde and U.~G.~Meissner,
  Phys.\ Rev.\ D {\bf 74} (2006) 034021
  [arXiv:hep-ph/0606133].

\bibitem{guagnelli}
M.~Guagnelli, J.~Heitger, R.~Sommer and H.~Wittig  [ALPHA Collaboration],
Nucl.\ Phys.\ B {\bf 560} (1999) 465
[arXiv:hep-lat/9903040].

\bibitem{source}
  P.~de Forcrand, G.~Schierholz, H.~Schneider and M.~Teper,
  Phys.\ Lett.\ B {\bf 152} (1985) 107.

\bibitem{jliu}
K. Jansen and C. Liu,
\newblock Comput. Phys. Commun. 99 (1997) 221, hep-lat/9603008.

\bibitem{hasenb}
M.~Hasenbusch,
Phys.\ Lett.\ B {\bf 519} (2001) 177
[arXiv:hep-lat/0107019];\\
M.~Hasenbusch and K.~Jansen,
Nucl.\ Phys.\ B {\bf 659} (2003) 299
[arXiv:hep-lat/0211042].

\bibitem{JLQCD}
JLQCD, S. Aoki et~al.,
\newblock Phys. Rev. D65 (2002) 094507, hep-lat/0112051.

\bibitem{hasenb-LAT05}
  M.~Hasenbusch,
  PoS {\bf LAT2005} (2006) 116
  [arXiv:hep-lat/0509080].


\bibitem{qcdsf}
A.~Ali Khan {\it et al.}  [QCDSF Collaboration],
Phys.\ Lett.\ B {\bf 564} (2003) 235
[arXiv:hep-lat/0303026].

\bibitem{urbach}
C.~Urbach, K.~Jansen, A.~Shindler and U.~Wenger,
Comput.\ Phys.\ Commun.\  {\bf 174}, 87 (2006)
[arXiv:hep-lat/0506011].

\bibitem{Sexton:1992nu}
J.~C.~Sexton and D.~H.~Weingarten,
Nucl.\ Phys.\ B {\bf 380} (1992) 665.

\bibitem{alpha-unpublished} ALPHA collaboration, unpublished.

\bibitem{sap3}
  M.~L\"uscher,
  Comput.\ Phys.\ Commun.\  {\bf 165} (2005) 199
  [arXiv:hep-lat/0409106].

\bibitem{luscher-LAT05}
  M.~L\"uscher,
  PoS {\bf LAT2005} (2006) 002
  [arXiv:hep-lat/0509152].




\bibitem{tlength}
  H.~B.~Meyer, H.~Simma, R.~Sommer, M.~Della Morte, O.~Witzel and U.~Wolff,
   Comput. Phys. Commun. (in print),   arXiv:hep-lat/0606004.
\end{thebibliography}
\end{document}